\documentclass[12pt]{article}

\usepackage{amsmath}
\usepackage{amsfonts}
\usepackage{amssymb}
\usepackage{graphicx}
\usepackage{fancyvrb}
\usepackage{graphicx}
\usepackage{hyperref}
\usepackage{fancyref}
\usepackage{mathtools}
\usepackage{comment}
\VerbatimFootnotes
\usepackage{a4wide}

\usepackage{caption}
\usepackage{subcaption}

\begin{document}

\begin{flushright}\footnotesize

\texttt{}
\vspace{0.6cm}
\end{flushright}

\mbox{}
\vspace{0truecm}
\linespread{1.1}

\centerline{\LARGE \bf Holographic Correlators of Giant Gravitons }
\medskip
\medskip
\centerline{\LARGE \bf in Monodromy Defects}
\medskip

\vspace{.5cm}

 \centerline{\LARGE \bf }

\vspace{1.5truecm}


\centerline{
    { \bf Diego Rodriguez-Gomez${}^{a,\,b,\,\dagger}$}}

\begingroup
\renewcommand{\thefootnote}{\textdagger}
\footnotetext{d.rodriguez.gomez@uniovi.es}
\endgroup

\vspace{1cm}
\centerline{{\it ${}^a$ Department of Physics, Universidad de Oviedo}} \centerline{{\it C/ Federico Garc\'ia Lorca  18, 33007  Oviedo, Spain}}
\medskip
\centerline{{\it ${}^b$  Instituto Universitario de Ciencias y Tecnolog\'ias Espaciales de Asturias (ICTEA)}}\centerline{{\it C/~de la Independencia 13, 33004 Oviedo, Spain.}}
\vspace{1cm}

\centerline{\bf ABSTRACT}
\medskip 

We compute holographically correlation functions for giant gravitons in $\mathcal{N}=4$ SYM in the presence of monodromy defects through probe branes. The computation boils down to the study of charged geodesics in certain five-dimensional gauged supergravity backgrounds. In addition to the standard U-shaped geodesic, in the presence of the defect, we find an extra, novel, contribution from a geodesic anchored at the defect which captures the one-point function of the square of the giant graviton.

\noindent 

\setcounter{footnote}{0}

\newpage

\tableofcontents

\section{Introduction}

Defect operators are very interesting objects in Quantum Field Theory (QFT). In particular, since they break in a controlled manner spacetime symmetries, they allow to probe finer details of the QFT. Moreover, as it has become clear over the last decade, topological defects within a theory provide a vantage approach to the symmetries of the system. In this paper we consider a class of codimension two defects defined by the monodromy that operators charged under a global internal $U(1)$ 0-form symmetry pick as they encircle the defect in the transverse plane. For obvious reasons, these defects are dubbed monodromy defects. In modern parlance, one can regard monodromy defects as the endpoints of the codimension one topological operators which implement the abelian 0-form symmetry; thus tying together the two aspects of defect operators raised above. From this perspective, as pointlike operators $O$ with charge $m$ under a $U(1)$ are transported around the monodromy defect in the transverse plane --let us denote the polar angle by $\theta$--, they must cross the symmetry operator and pick up the phase which defines the monodromy operator as in \eqref{monodromyoperator}

\begin{equation}
\label{monodromyoperator}
O(\theta+2\pi)=e^{-i\,2\,\pi\,m\,\beta}\,O(\theta)\,.
\end{equation}

In this paper we are interested in monodromy defects for the R-symmetry in $\mathcal{N}=4$ SYM with unitary gauge group of rank $N$, focusing in particular on their holographic description. As monodromy defects can be regarded as the endpoint of symmetry defects, an additional motivation to study these holographically is that they may potentially allow to test further properties of the recently introduced --albeit in the context of $\mathcal{N}=1$ 4d SCFT's-- holographic R-symmetry defects \cite{Calvo:2025usj}.\footnote{See also \cite{Bah:2025vfu}.}

Coming back to the set-up of interest in this paper, the QFT has an $SO(6)$ R-symmetry which has a $U(1)^3$ Cartan torus under which the three complexified scalars are charged. Thus, we can imagine a monodromy defect as defined above for each of the three $U(1)$'s (and their arbitrary combinations, of course). For definiteness, let us consider a monodromy defect along the specific $U(1)_1$ inside the Cartan, under which the complex scalar $\phi_1$ is charged. Natural quantities to compute are two-point functions of charged operators in the presence of the defect. This leads us to consider operators of the schematic form $O=\phi_1^m$. For $m$ of $\mathcal{O}(1)$ we expect $O$ to holographically correspond to a KK fluctuation of the dual gravitational background corresponding to the defect inserted in $\mathcal{N}=4$ SYM. In turn, for $m$ of $\mathcal{O}(N)$ we expect $O$ to holographically correspond to a probe brane in that geometry. A particular example of the latter class are giant gravitons, which are $\frac{1}{2}$ BPS operators with dimension $\Delta=m$ and $U(1)_1$ R-charge $Q=m$, corresponding to subdeterminants in QFT \cite{Balasubramanian:2001nh} and represented by D3 branes moving in the internal space  \cite{McGreevy:2000cw,Grisaru:2000zn}. Motivated by this, as probe branes are much easier to study, we will study two-point functions for giant gravitons in $\mathcal{N}=4$ SYM. To be more specific, we consider the QFT on $\mathbb{R}^4$ with metric

\begin{equation}
ds^2=d\vec{\sigma}^2+d\vec{x}^2\,,
\end{equation}
and place the monodromy defect along $\vec{\sigma}$ at the origin of $\vec{x}$. We will then holographically study correlation functions of giant gravitons in the presence of the monodromy defect in the simplified kinematical regime where we insert the giant gravitons at $\vec{x}=(\pm X,Y)$ and same $\vec{\sigma}$, so that the correlation function we are interested in is $\langle O(x_1) O^{\dagger}(x_2)\rangle$, with

\begin{equation}
x_1=(x_1^{\parallel}=\vec{\sigma},\,x_1^{\perp}=(X,Y)))\,,\qquad x_2=(x_2^{\parallel}=\vec{\sigma},\,x_2^{\perp}=(-X,Y))\,.
\end{equation}
%
%
While this is a particular kinematical regime --which corresponds to $\chi=2$ in the notation of  \cite{Billo:2016cpy}-- and hence will not allow us to find the two-point function in full generality, it will be enough to allow us to extract in a safe manner the one-point function for the operator $OO^{\dagger}$.

It remains to discuss the appropriate set-up to find the gravitational dual to $\mathcal{N}=4$ SYM with the insertion of a monodromy defect. Since the defect is characterized by the holonomy of the $U(1)^3$ inside the R-symmetry as we go around, it seems natural to guess that the minimal set-up to capture the configuration of interest is the $U(1)^3$ truncation of 5d $SO(6)$ gauged SUGRA --the so-called STU model-- whose Type IIB uplift was discussed in \cite{Cvetic:2000nc}. In fact, the relevant solutions to the 5d SUGRA have been constructed in \cite{Kunduri:2007qy,Ferrero:2021etw} and recently studied in \textit{e.g.} \cite{Arav:2024exg,Bomans:2024vii,Conti:2025wyj}. As shown in \cite{Bomans:2024vii}, when appropriately choosing the parameters, these solutions coincide with a class of the bubbling solutions of \cite{Gomis:2007fi} also recently studied in \cite{Choi:2024ktc,IzquierdoGarcia:2025jyb}. The latter are the holographic duals to the surface operators introduced in \cite{Gukov:2006jk}, and generically involve, in addition to monodromy, non-trivial extra dynamical ingredients of the ambient $\mathcal{N}=4$ SYM (to begin with, they require the choice of a Levi subgroup within gauge group). As we are interested in the much simpler monodromy defects defined above, we will turn off these extra degrees of freedom, so that our monodromy defects, strictly speaking, lie outside of the classes studied in these references.

As we will show, the problem of studying holographic correlators for giant gravitons boils down to studying charged geodesics hanging from the boundary in the 5d gauged SUGRA. This is essentially the WKB approximation to the two-point function \cite{Janik:2010gc} (see \cite{Bak:2011yy,Bissi:2011dc} for applications to a context similar to ours). It turns out that, in the presence of the monodromy defect, there are two relevant such geodesics. One is the direct relative of the U-shaped geodesic capturing the two-point function in the absence of the defect. The other is a novel saddle corresponding to a geodesic anchored to the defect whose existence is tied to the presence of the defect. The full two-point function is the sum of both contributions, with the anchored contribution capturing the one-point function for the operator $OO^{\dagger}$ in the presence of the defect. It is important to emphasize that while the particular coefficients will depend on details of the defect, the structure and presence of the contribution itself is universal, only controlled by the existence of the geodesic and the asymptotics of the background.

The outline of the paper is as follows. In Section \ref{sec:geodesics} we study giant gravitons in the uplifted STU model further truncated to a single scalar. It turns out that while originally giant gravitons correspond to D3 branes wrapping an internal $S^3$, the problem in the end boils down to studying charged geodesics in the 5d gauged SUGRA. In Section \ref{sec:background} we introduce and analyze explicitly the background of interest. In particular, we show how indeed giant gravitons pick the expected phase for the corresponding monodromy defect and discuss the turning-off of the degrees of freedom living at the defect. In Section \ref{sec:correlationfunctions} we compute the two-point function of giant gravitons through charged U-shaped geodesics. In addition to the standard U-shaped geodesic, in the presence of the defect there is another saddle corresponding to a geodesic anchored at the defect, so that the two-point function is the sum of these two contributions. It turns out that the anchored geodesic kicks-in abruptly as $\beta_i$ is turned on. In Section \ref{sec:anchored?} we discuss possible physical mechanisms smoothing this transition. We conclude in Section \ref{sec:conclusions} with an outlook and final thoughts, including some speculations about the putative relation to the symmetry operators introduced in  \cite{Calvo:2025usj}. We leave to Appendix \ref{app:analyticity} some technical details of the evaluation of the on-shell action for the geodesics. 

\section{Giant gravitons as charged geodesics in 5d gauged supergravity}
\label{sec:geodesics}

Let us consider 5d $SO(6)$ gauged supergravity upon the further truncation to the $U(1)^3$ invariant sector --the often called STU model. This keeps three $U(1)$ gauge fields $A_1=A_{12}$, $A_2=A_{34}$ and $A_3=A_{56}$ as well as the diagonal scalar matrix

\begin{equation}
T={\rm diag}(X_1,\,X_1,\,X_2,\,X_2,\,X_3,\,X_3)\,,\qquad X_1\,X_2\,X_3=1\,.
\end{equation}
Solutions to this model can be uplifted to 10d Type IIB SUGRA using the formulae in \cite{Cvetic:2000nc}. We will use the parametrization for the internal space

\begin{equation}
\label{eq:mus}
\begin{array}{lll}
\mu_1=\cos\psi\,\cos\phi\,, & & \mu_2=\cos\psi\,\sin\phi\,,\\
\mu_3=\sin\psi\,\cos\frac{\alpha_1}{2}\,\cos\frac{\alpha_3+\alpha_2}{2}\,,& & \mu_4=\sin\psi\,\cos\frac{\alpha_1}{2}\,\sin\frac{\alpha_3+\alpha_2}{2}\,,\\
\mu_5=\sin\psi\,\sin\frac{\alpha_1}{2}\,\cos\frac{\alpha_3-\alpha_2}{2}\,,& & \mu_6=\sin\psi\,\sin\frac{\alpha_1}{2}\,\sin\frac{\alpha_3-\alpha_2}{2}\,.
\end{array}
\end{equation}

As an appetizer, let us take the further restriction to the case where $X_1=X_2=X_3=1$. We now consider a D3 brane wrapping the $S^3$ parametrized by $\{\alpha_1,\alpha_2,\alpha_3\}$ and a curve --parametrized by $\xi$-- in the 5d space. After a short computation, it is straightforward to see that the action for such a brane is 

\begin{equation}
S_{D3}=\int d\xi\, 2\pi^2\,T_3\,\sin^3\psi\,\sqrt{g_5+\cos^2\psi\,(\phi'-g\,A_1)^2}-i\,2\pi^2\,T_3\,\sin^4\psi\,(\phi'-A_1)\,;
\end{equation}
where prime denotes derivative with respect to $\xi$ and $g_5$ and $A_1$ stands for the pull-back of the 5d metric and gauge fields to the 1d curve parametrized by $\xi$. Eliminating the cyclic coordinate $\phi$ into a routhian one finds\footnote{Note that so far we are being agnostic about the signature of the space, entirely encoded in $g_5$. However, the computation looks formally euclidean, and hence the $i$ in front of the momentum.}

\begin{equation}
\mathcal{R}=-\sqrt{g_5}\, 2\pi^2\,T_3\,\sin^3\psi\,\sqrt{1+\Big(1-i\frac{m}{2\pi^2T_3\sin^4\psi}\Big)^2\,\tan^2\psi}+m\,A_1\,,
\end{equation}
where $m$ is the conserved momentum. Extremizing with respect to $\psi$ fixes

\begin{equation}
\label{J}
m=-i\,2\pi^2\,T_3\,\sin^2\psi\,,
\end{equation}
so that the on-shell routhian is

\begin{equation}
\mathcal{R}=-i\,m\,(\sqrt{g_5}+i\,A_1)\,.
\end{equation}
Thus we see that effectively our D3 brane behaves like a charged particle, with charge equals to the mass, in the 5d geometry, precisely as expected for a giant graviton.\footnote{We are assuming the appropriate boundary terms so as to impose fixed momentum. See \cite{Yang:2021kot,Holguin:2025dei} for a more thorough discussion.} The mass $m$ coincides with the dimension of the giant graviton and charge equaling the mass reflects the BPS condition saturated by the giant. We emphasize that \eqref{J} shows that $m\sim N$.

Consider now the more general case where the scalar matrix is

\begin{equation}
T={\rm diag}(\mathcal{X}^{-2},\,\mathcal{X}^{-2},\,\mathcal{X},\,\mathcal{X},\,\mathcal{X},\,\mathcal{X})\,.
\end{equation}
The computation is now much more cumbersome, but the final result is that the $\psi$ equation of motion is solved by the same value of $P$ as in \eqref{J} and

\begin{equation}
\mathcal{R}=-im\,(\mathcal{X}^{-2}\,\sqrt{g_5}+i\,A_1)\,.
\end{equation}
Thus the final upshot is that giant gravitons end up behaving like massive charged particles in 5d gauged SUGRA backgrounds. 

\section{Monodromy defects  in 5d gauged SUGRA}\label{sec:background}

The backgrounds of 5d $d=5$ $U(1)^3$ gauged SUGRA of interest for us were first constructed in \cite{Kunduri:2007qy,Ferrero:2021etw} and recently studied in \textit{e.g.} \cite{Arav:2024exg,Bomans:2024vii,Conti:2025wyj}. Setting for simplicity $\nu=0$ (and $g=1$) these backgrounds are 

\begin{align}
\label{eq:background}
& ds^2=\frac{H^{\frac{1}{3}}}{g^2}\,\Big(ds_{AdS_3}^2+\frac{1}{4P}dy^2+\frac{P}{H}\,d\theta^2\Big)\,,\nonumber \\
& A_I=\Big(\alpha_I-\frac{y}{y+q_I}\Big)\,d\theta\,, \qquad X_I=\frac{H^{\frac{1}{3}} }{(y+q_I)} \,; \nonumber \\
& H=(y+q_1)(y+q_2)(y+q_3)\,;
\end{align}
where $P=H-y^2$. Obviously the background makes sense up to some minimal $y_{\star}$, where the circle $\theta$ collapses. Close to $y_{\star}$, we set $y=y_{\star}+\rho^2$, so for small $\rho$ the metric is

\begin{equation}
ds^2=\frac{H_{\star}^{\frac{1}{3}}}{g^2}\,ds_{AdS_3}^2+\frac{H_{\star}^{\frac{1}{3}}}{g^2}\,\frac{1}{P'_{\star}}\Big(d\rho^2+T^{-2}\rho^2\,d\theta^2\Big)\,,\qquad T=|\frac{\sqrt{H_{\star}}}{P'_{\star}}|\,;
\end{equation}
where $_{\star}$ means evaluation at $y_{\star}$.

Let us define the monodromies at $y_{\star}$ as $A^I=h_I\,d\theta$, and the monodromies at $y\rightarrow\infty$ as $A^I=-\beta_I\,d\theta$. To begin with, we immediately have $\alpha_I=-\beta_I+1$. Moreover, it is straightforward to see that

\begin{equation}
y_{\star}=\big(1-(h_1+\beta_1)\big)\,\big(1-(h_2+\beta_2)\big)\, \big(1-(h_3+\beta_3)\big)\,.
\end{equation}
And

\begin{align}
& q_1=(h_1+\beta_1)\,\big(1-(h_2+\beta_2)\big)\,\big(1-(h_3+\beta_3)\big)\,, \\
& q_2=(h_2+\beta_2)\,\big(1-(h_1+\beta_1)\big)\,\big(1-(h_3+\beta_3)\big)\,,\\ 
& q_3=(h_3+\beta_3)\,\big(1-(h_1+\beta_1)\big)\,\big(1-(h_2+\beta_2)\big)\,.
\end{align}
Finally

\begin{equation}
T=|\frac{1}{1-(h_1+\beta_1)-(h_2+\beta_2)-(h_3+\beta_3)}|\,.
\end{equation}
Thus we see that unless $h_I+\beta_I=0$ --which makes all $q_I=0$--, $T\ne 1$ and correspondingly the background will have a singularity at $y_{\star}$.

The giant gravitons discussed above correspond to charged geodesics in this background provided we set $q_2=q_3$ (and consequently $\beta_2=\beta_3$) and $X_1=\mathcal{X}^{-2}$. For concreteness, the identification of parameters is

\begin{equation}
y_{\star}=(1-\beta_1)\,(1-\beta_2)^2\,,\qquad q_1=\beta_1\,(1-\beta_2)^2\,, \quad  q_2=\beta_2\,(1-\beta_2)\,(1-\beta_1)\,.
\end{equation}

\subsection{The region asymptotically far away from the defect}

Let us consider the large $y$ asymptotics. The metric goes to

\begin{equation}
ds^2=y\,\Big(ds_{AdS_3}^2+\frac{1}{4y^3}dy^2+d\theta^2\Big)\,.
\end{equation}
Writing $ds^2_{AdS_3}=u^{-2}\,(d\vec{\sigma}^2+du^2)$, it is straightforward to check that upon doing

\begin{equation}
\label{change}
(u,\,y,\,\tan\theta)\rightarrow (\sqrt{\vec{x}^2+z^2},\,\frac{\vec{x}^2+z^2}{z^2},\,\frac{x_2}{x_1})\,,
\end{equation}
the metric becomes that of $AdS_5$ with radial coordinate $z$ and boundary $\{\vec{\sigma},\vec{x})$. Note that $z\rightarrow 0$ corresponds to $y\rightarrow \infty$. In that region the gauge fields are

\begin{equation}
A_I\sim -\beta_I\,d\theta\,.
\end{equation}
Thus fixing $\beta_I$ amounts to fixing a monodromy defect for the R-symmetry supported at $\vec{x}=0$ within $\mathcal{N}=4$ SYM. 

Let us consider a giant graviton at very large and fixed $y$ wrapping $t$ in Poincare coordinates and with $\theta=\theta(t)$. Using our charged geodesic recipe the action for such brane is (we use \eqref{change} to write things in terms of $AdS_5$ coordinates)

\begin{equation}
S = \int \frac{m}{z}\,\sqrt{1+\dot{\theta}^2} - m\beta_1\dot{\theta}
\end{equation}
The canonically conjugated momentum to $\dot{\theta}$ --call it $p_{\theta}$-- is conserved and

\begin{equation}
\dot{\theta}=\frac{z\,(p_{\theta}+m\beta_1)}{\sqrt{m^2-z^2\,(p_{\theta}+m\beta_1)^2}}\,.
\end{equation}
In order to have a mass $m$ particle --as it should be for a giant graviton-- we must set $p_{\theta}=-m\beta_1$. Thus, the giant graviton wavefunction $e^{-ip_{\theta}\theta}$ makes it to pick a phase $2\pi m\beta_1$ as it $\theta\rightarrow \theta+2\pi$, in particular showing that the defect is indeed a monodromy defect. Let us stress that our choice of operator --a D3 spinning along $\phi$ corresponding to an operator $O\sim \phi_1^m$-- only probes the $U(1)_1$ component of the monodromy operator --that is, the monodromy is blind to $\beta_2$.

\subsection{On the monodromy at the origin}
 
Let us write $AdS_3$ in global coordinates and consider a D3 brane along $t$ at the center of global $AdS_3$ with $\theta=\theta(t)$ sitting at $y_{\star}$. We expect this brane to represent a giant graviton of $\mathcal{N}=4$ being brought on top of the defect. The action for such a brane, which remains well-defined even though it sits at $y_{\star}$, is

\begin{equation}
\label{eq:defectdof}
S=\int dt \,m \,\big(1-(h_1+\beta_1)\big)^{\frac{2}{3}}\,\big(1-(h_2+\beta_2)\big)^{\frac{1}{3}}+m\,h_1\,\dot{\theta}
\end{equation}
Note that for $h_1\ne 0$, $\theta$ appears as a topological zero mode of the operator, which therefore comes in a whole degenerate family (an infinite tower of operators with the same dimension). This signals the existence of a Hilbert space attached to the defect operator\footnote{The existence of the zero mode is purely topological, it is not fixed whatsoever from bulk data. It is thus reasonable to think that this extra degrees of freedom are solely localized in the defect. Upon quantization, for $h_1\ne 0$, it leads to a $\mathbb{C}$ Hilbert space attached to the operator.}, which suggests that the defect carries intrinsic dynamical degrees of freedom as it is familiar from other cases such as Gukov-Witten surface operators. In the following we will choose to study rigid defect operators and fix $h_I=0$, thereby eliminating such intrinsic defect sector and concentrating in defects solely defined by their monodromy.

Note now that $T$ is completely given in terms of $h_I+\beta_I$. Thus, fixing $T$ --perhaps quantized-- amounts to fixing the combinations $h_I+\beta_I$. Since we are setting $h_I=0$ while at the same time allowing arbitrary $\beta_I$, our $T$ is not fixed, nor quantized. This is in contrast to \cite{Bomans:2024vii}, where it was argued that the 5d STU solutions, upon appropriately tuning the parameters, can be mapped to the bubbling geometries of \cite{Gomis:2007fi} dual to Gukov-Witten defect operators \cite{Gukov:2006jk}.\footnote{The change of coordinates in Section 4.3 of \cite{Bomans:2024vii} maps the solutions with $q_2=q_3=0$ and $\beta_I=0$ to bubbling solutions of  \cite{Gomis:2007fi}, identifying $T$ with the rank of the Levi subgroup $k$. On the contrary, we are fixing $h_I=0$ and leaving $\beta_I$ free.} This is not a contradiction, as Gukov-Witten operators involve additional structural data (\textit{e.g.} a Levi subgroup choice) and are often naturally described as coupled to an intrinsic defect sector. This is explicitly reflected in the appearance of the topological zero mode $\theta$ for $h_I\ne 0$. By contrast, we consider defects specified solely by an R-symmetry monodromy, and hence set $h_I=0$. Note that such defects can still host a nontrivial defect CFT induced by the ambient theory, but we stress that we do not introduce or tune any additional independent defect degrees of freedom beyond the monodromy parameters. 

%
%

\section{Correlation functions from charged geodesics}\label{sec:correlationfunctions}

We now turn to the computation of two-point functions for giant gravitons in the presence of the monodromy defect. Since this is most conveniently done in euclidean signature, our background will be the euclidean version of \eqref{eq:background} (to be specific, we consider Poincare $AdS_3$ in euclidean signature). In this background we consider now a particle of charge $m$ and mass $m\,\mathcal{X}^{-2}$. We will be interested in geodesics departing/ending the $AdS_5$ boundary points $\{\vec{\sigma},\pm X,Y\}$ (that is, at large $y$). Using \eqref{change} we can write

\begin{equation}
y_{\rm M}=\frac{X^2+Y^2}{\epsilon^2}\,,\qquad \tan\theta=\frac{Y}{X}\,,\qquad u=\sqrt{X^2+Y^2}\,.
\end{equation}
and we have added a subscript $_{\rm M}$ to $y$ to denote that this is the maximal, regulated, value of $y$. The regulator $\epsilon$ is such that the $AdS_5$ boundary is at $z=\epsilon$.

Assuming $\theta=\theta(y)$, the action is 

\begin{equation}
S=\int m\,\mathcal{X}^{-2}\,H^{\frac{1}{6}}\,\sqrt{\frac{P}{H}(\theta')^2+\frac{1}{4P}}+i\,m\,A\,\theta'\,,
\end{equation}
where $A=\Big(\alpha-\frac{y}{y+q}\Big)$. Wick-rotating as well the CS term

\begin{equation}
S=\int dy\, f_1\,\sqrt{1+f_2\,(\theta')^2}+f_3\,\theta'\,;\qquad f_1=m\frac{\mathcal{X}^{-2}\,H^{\frac{1}{6}}}{2P^{\frac{1}{2}}}\,,\qquad f_2=\frac{4P^2}{H}\,,\qquad f_3=m\,A\,.
\end{equation}
From here we can find a first order equation of motion

\begin{equation}
\label{eq:eom}
\theta'=\frac{(J-f_3)}{\sqrt{f_2}\,\sqrt{f_1^2\,f_2-(J-f_3)^2}}\,.
\end{equation}
For this to make sense $f_1^2f_2>(J-f_3)$. Thus, this equation defines a minimal $y_{\rm m}$. In turn, the on-shell action is

\begin{equation}
\label{os}
S=\int_{y_{\rm m}}^{y_M}dy\,\frac{f_1^2\,f_2+f_3\,(J-f_3)}{\sqrt{f_2}\,\sqrt{f_1^2\,f_2-(J-f_3)^2}}\,.
\end{equation}

\subsection{Warm-up: removing the defect}

Upon removing the defect (that is, setting the $q$'s to zero) the minimal $y$ condition sets

\begin{equation}
y_{\rm m}=1+\frac{J_0^2}{m^2}\,.
\end{equation}
Moreover, the equation of motion can be easily integrated into

\begin{equation}
\tan(\theta-\theta_0)=-\frac{m+\sqrt{m^2(y-1)-J_0^2}\,\sqrt{y}-m\,y}{J_0}\,.
\end{equation}
At the minimal $y_{\rm m}$ we expect $\theta_{\rm m}=\frac{\pi}{2}$. This fixes  

\begin{equation}
\tan\theta_0=\frac{m}{J_0}\,.
\end{equation}
In turn, for large $y$, where $\tan\theta=\frac{Y}{X}$, we have 

\begin{equation}
\tan(\theta-\theta_0)=\frac{J_0}{2m}-\frac{m}{2J_0}\qquad \leadsto \qquad \frac{J_0}{m}=\frac{Y}{X}
\end{equation}
Note that as a consequence, for large $y$

\begin{equation}
\label{eq:as}
\theta'=-\frac{J_0}{2\,m\,y^2}+\cdots\,.
\end{equation}
Plugging all this into the action it evaluates to 

\begin{equation}
\label{halfU}
S_{\rm os}=m\log\frac{2X}{\epsilon}\,,
\end{equation}
so that the giant graviton two-point function is

\begin{equation}
\langle O(-X,Y)\,O^{\dagger}(X,Y)\rangle = e^{-2S_{\rm os}}=\frac{\epsilon^{2m}}{(2X)^{2m}}\,.
\end{equation}
As expected, in absence of the monodromy defect the $Y$ dependence of the correlation function drops, and it only depends on the distance between the insertions.

\subsection{Correlation functions in the presence of the defect}

We now come back to the case of interest where the defect is present. It is natural to expect that in the more complicated background including the defect an analog of the U-shaped geodesics of the no-defect-case will be present. We will refer to these as standard geodesics. However, since the background has a singularity at $y_{\star}$, one can imagine geodesics anchored at the singularity, running from the boundary to $y_{\star}$. We will dub these anchored geodesics. Given the existence of two saddles, the full correlation function will be the sum of the exponentiated action of each.

\subsubsection{Standard geodesics}

Note first that for large $y$, from the equation of motion \eqref{eq:eom} we have

\begin{equation}
\theta'=\frac{J+m\,\beta_1}{2my^2}+\cdots \,.
\end{equation}
Since the background is asymptotically $AdS$, one natural class of geodesics is that with the same asymptotics \eqref{eq:as} as in the case without the defect

\begin{equation}
J=-J_0- m\,\beta_1\,.
\end{equation}
We should now plug this into \eqref{os} and perform the corresponding integral. Unfortunately, an analytic treatment seems impossible. Thus, in order to further proceed, let us assume that $\beta_1\sim\beta_2$ and perform a perturbative expansion in $\beta_I$. Then, as for the turning point one finds

\begin{equation}
\label{eq:ymstandard}
y_{\rm m}=(1+\frac{J_0^2}{m^2})+\frac{(2mJ_0+J_0^2-m^2)\,\beta_1-2\,(J_0^2+m^2)\,\beta_2}{J_0^2+m^2}+\cdots\,.
\end{equation}
Using this, one can compute the on-shell action (see appendix \ref{app:analyticity} for relevant technical details), finding

\begin{equation}
e^{-2S_{\rm os}^{\rm(standard)}}=\frac{\epsilon^{2m}}{(2X)^{2m}}\,\Big[1 + \frac{2\,m\,X}{(X^2+Y^2)^2}\,\Big( (Y^3-X^2Y-2XY^2)\,\beta_1+2(X^3+XY^2)\,\beta_2\Big) +\cdots \Big]
\end{equation}

\subsubsection{Anchored geodesics}

As anticipated, we can have another class of geodesics going all the way to $y_{\star}$ (\textit{i.e.} for these $y_{\rm m}=y_{\star}$) where the circle shrinks at $P=0$. Since $f_1^2f_2\sim P$, this can be enforced by setting

\begin{equation}
J=f_3^{\star}\,,
\end{equation}
where $^{\star}$ means evaluation at $y_{\star}$. Note that since we set $h_I=0$, $f_3^{\star}=0$, so that $J=0$. An important consequence is that at $y_{\star}$ the boundary term of the variational problem vanishes, so that indeed no extra boundary term is needed --in accordance with our observation that $h_I=0$ turns off defect degrees of freedom.

For this geodesic the spacetime dependence is very easy to compute: since $y_{\rm m}=y_{\star}$ is independent of $X,\,Y$, all the $X,\,Y$ dependence will come from $y_M$. Thus, we can simply expand for large $y$ \eqref{os}, integrate and evaluate at $y_M$ to find that this geodesic contributes $\sim m\,\log \frac{4\sqrt{X^2+Y^2}}{\epsilon}$. In order to find the precise coefficient we should performing the integral, finding

\begin{equation}
\label{eq:Sanchored}
e^{-2S_{\rm os}^{\rm (anchored)}}=\frac{\epsilon^{2m}}{(2X)^{2m}}\,\frac{e^{-4\,m\,\beta_2}}{(1+\frac{Y^2}{X^2})^{m}}\,.
\end{equation}

Note that we have chosen to keep the exponential form (even though we are taking the $\beta_I\rightarrow 0$ limit). More importantly, note that \eqref{eq:Sanchored} does not vanish in the limit $\beta_I\rightarrow 0$. This simply reflects that even in smooth case the distance from the boundary to $y_{\star}$ is non vanishing. However, as a configuration describing a D3 brane, this geodesic ceases to be meaningful if $\beta_1=0$. Thus, \eqref{eq:Sanchored} implicitly comes multiplied by a Heaviside function $\Theta(\beta_1)$. In Section \ref{sec:anchored?} we discuss a possible interpretation of this $\Theta(\beta_1)$.

\subsection{The full two-point function}

Putting all pieces together, our prediction for the two-point function for giant gravitons in the presence of a monodromy defect is

\begin{align}
\label{eq:two-point}
& \langle O(X,Y)\,O^{\dagger}(-X,Y)\rangle=\\ &  \frac{\epsilon^{2m}}{(2X)^{2m}}\,\Big[1 + \frac{2\,m\,X}{(X^2+Y^2)^2}\,\Big( (Y^3-X^2Y-2XY^2)\,\beta_1+2(X^3+XY^2)\,\beta_2\Big)  + \Theta(\beta_1)\,\frac{e^{-4m\beta_2}}{(1+\frac{Y^2}{X^2})^m}+\cdots \Big]
\end{align}

This result has precisely the expected structure for a two-point function \cite{Billo:2016cpy}. In particular, it allows to read-off the contribution of $OO^{\dagger}$ operator by taking the $X\rightarrow 0$ limit \cite{Bianchi:2021snj}. Subtracting the obvious contact divergence, the one-point function of $OO^{\dagger}$ in the presence of the defect is

\begin{equation}
\langle OO^{\dagger}(Y)\rangle=\frac{e^{-4\,m\,\beta_2}}{2^m}\frac{\epsilon^{2m}}{Y^{2m}}\,.
\end{equation}
Note that this entirely comes from the anchored geodesic. In fact, looking at \eqref{eq:ymstandard} we see that taking $X\rightarrow 0$ at fixed $Y$ sends $y_{\rm m}$ to the boundary. Hence, in the limit which should produce the $Y^{-2m}$ term associated to the one-point function of $OO^{\dagger}$, the standard geodesic probes less and less the defect. On the other hand, the anchored geodesic remains, by definition, anchored; and produces as shown the one-point function. We stress that, since the spacetime dependence of the anchored geodesic entirely comes from the large $y$ region, while the actual coefficient of the one-point function might heavily depend on the details of the background, the fact that in the $X\rightarrow 0$ limit at fixed $Y$ the anchored geodesic will provide a term $\sim Y^{-2m}$ in the two-point function is guaranteed.

Coming back to the full expression \eqref{eq:two-point}, it would be very interesting to identify the exchanged operators performing a conformal block decomposition. Unfortunately, with our specific kinematical choice $\chi=2$, infinitely many blocks are masked together making the block decomposition very hard. We postpone for further work the computation of the two-point function with a more generic insertion of the operators which would allow to unambiguously identify the blocks.

\section{Comments on the anchored geodesics}\label{sec:anchored?}

Our computation for the anchored geodesic in \eqref{eq:Sanchored} captures the proper length from the boundary to the point $y_{\star}$ and consequently does not vanish when $\beta_1\rightarrow 0$. In order to understand this, let us pretend that anchored geodesics existed in the $\beta_1=0$ $AdS_5$ case. From \eqref{eq:eom}, we would have $\theta'=0$. With no loss of generality, we could take $\theta=0$ and correspondingly $Y=0$: we would be computing a $U$-shaped geodesic running along the $x$ axis, and the anchoring would be simply the point where the corresponding arm of the $U$ ends (and meets the other arm). In this case we would have $y_{\rm m}=1$ and

\begin{equation}
S=\int_1^{y_M}dy\,f_1=\frac{1}{2}m\,\log 4y_M=m\,\log\frac{X}{m}\,,
\end{equation}
which is precisely \eqref{halfU}. 

\begin{figure}[h!]
\centering
\includegraphics[scale=.17]{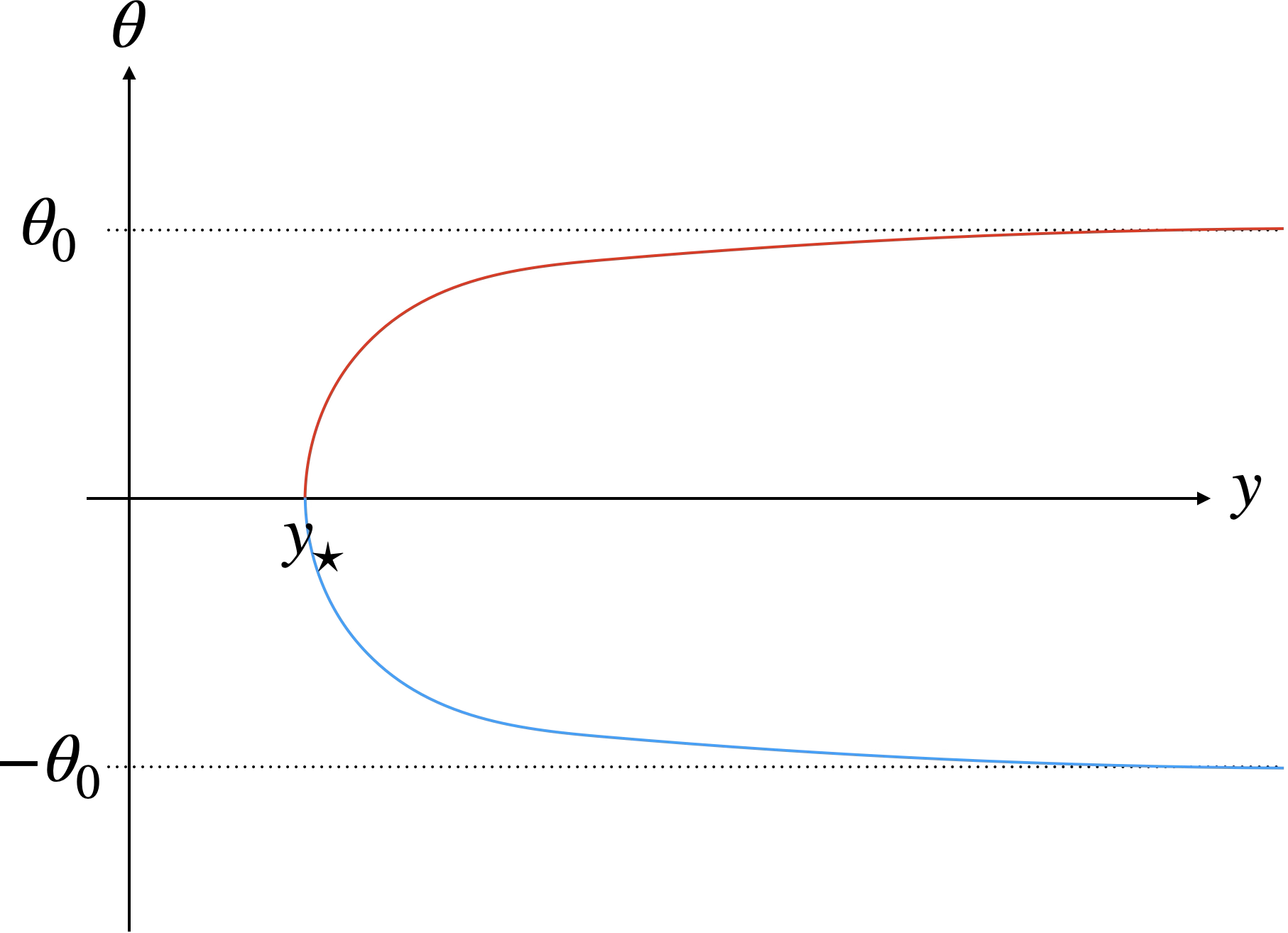}
\caption{Cartoon of the anchored geodesic.}
\label{fig:anchored}
\end{figure}

Coming back to the $\beta_1\ne 0$ case, from \eqref{eq:eom} we have

\begin{equation}
\begin{cases}y\sim y_{\star}:\quad 
\theta'=\frac{\beta_1}{2}\,\frac{1}{\sqrt{y-y_{\star}}}\,. \\ 
y\rightarrow \infty:\qquad \theta'=\frac{\beta_1}{2}\,\frac{1}{y^2}\,.
\end{cases}
\end{equation}
It should be stressed that in the case of the anchored geodesic, this result does not rely on a perturbative expansion in $\beta_1$. From here it follows that, provided $\beta_1\ne 0$, the anchored geodesic looks like the cartoon in \fref{fig:anchored}.
Thus we see that this is really a U-shaped geodesic touching $y_{\star}$ and interpolating from $\theta_0\leftrightarrow -\theta_0$ at large $y$. Since  $\tan\theta_0=\frac{Y}{X}$, this translates into $X$ to $-X$. It is now clear that if $\beta_1=0$ this geodesic simply does not exist.\footnote{To be more precise: it has to be interpreted as one of the arms of a regular U-shaped geodesic which, as it has $Y=0$, simply happens to pass through $y_{\star}$.} Thus, as described above, we expect that the contribution of the anchored geodesic comes multiplied by a $\Theta(\beta_1)$ factor. 

On the other hand, this non-analytic behavior in a two-point function is strange from the standpoint of finite $N$ QFT.\footnote{Note that the one-point function of square of the charged operator is not obviously controlled by a central charge and \textit{a priori} depends on the continuous defect parameters. For instance, in a similar situation albeit for a free theory, the computation of \cite{Bianchi:2021snj} shows a smooth turning-off of the 1-point function for the square of the charged operator.} Thus, it is natural to expect the existence of a mechanism to smooth this $\Theta$ behavior. One possible such mechanism is suggested by inspection of \fref{fig:anchored}, since, as $\beta_1\rightarrow 0$ the configuration looks like a brane/antibrane prone to decay through nucleation of D3-brane tubes breaking the anchored geodesic. A cartoon of the configuration would be as in \fref{fig:instability} below.

\begin{figure}[h!]
\centering
\includegraphics[scale=.17]{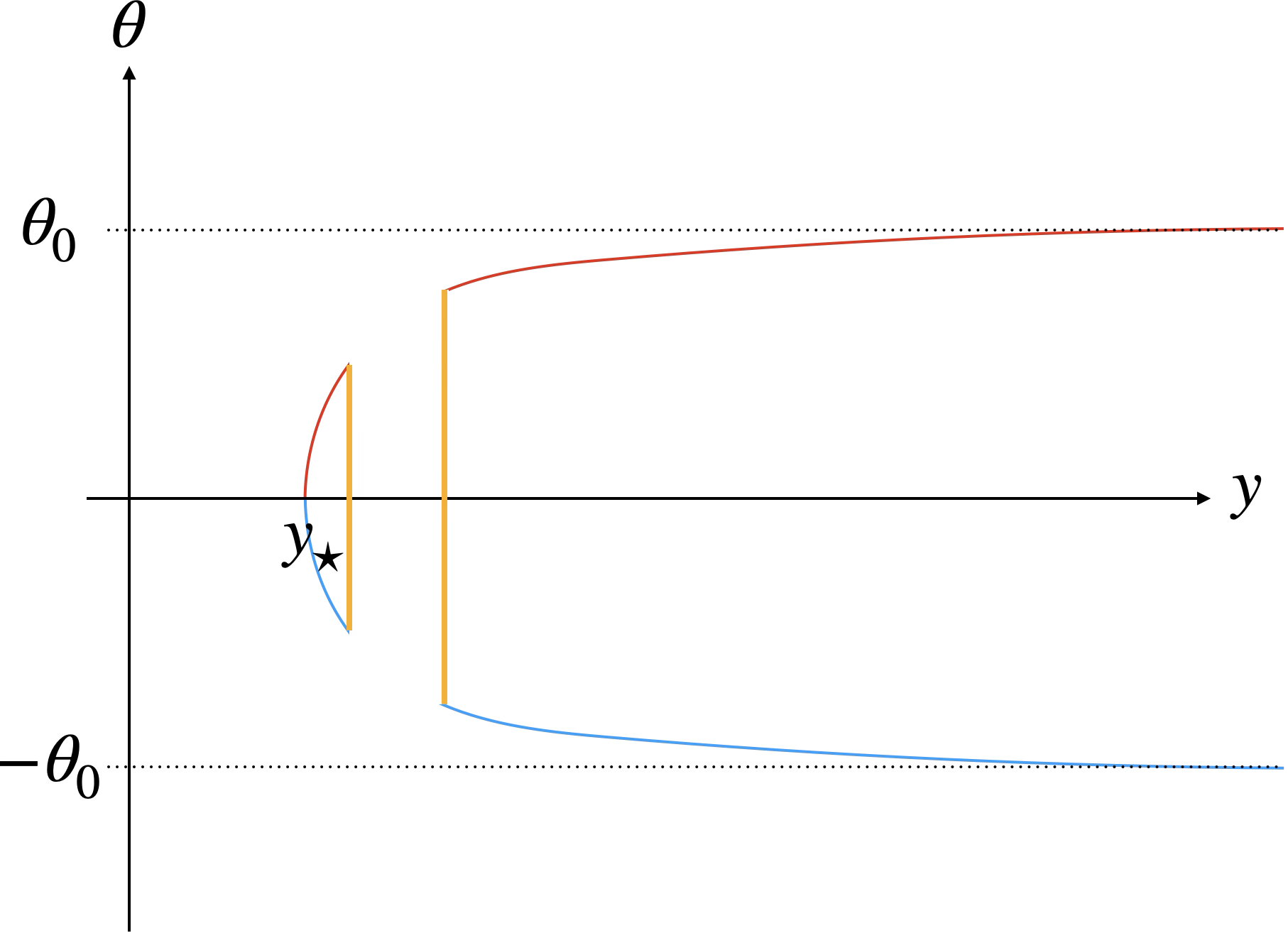}
\caption{Cartoon of the instability. The orange tube is the bubbled D3 which breaks the brane.}
\label{fig:instability}
\end{figure}
In order to estimate these instabilities, imagine a D3 tube going along $\theta$ from $-\theta_0$ to $\theta_0$ at fixed $y$. Its action is simply

\begin{equation}
S_{\rm D3-tube}=m\frac{P^{\frac{1}{2}}}{H^{\frac{1}{3}}}\,\Delta\theta\,.
\end{equation}
Due to the $y$-dependence of $P$ and $H$, the tube wants to fall towards $y_{\star}$. Around that region, for $y\sim y_{\star}$, we can approximate the action by

\begin{equation}
\label{eq:tube}
S_{\rm D3-tube}=m\frac{(P'_{\star})^{\frac{1}{2}}}{H_{\star}^{\frac{1}{3}}}\,\sqrt{y-y_{\star}}\,\Delta\theta\sim m\,\sqrt{y-y_{\star}}\,\Delta\theta\,.
\end{equation}
Since we want the tube to meet the two arms of the $U$, the $\Delta\theta$ should match that of the geodesic. In turn, using the equation of motion, we can estimate that for those $\Delta\theta \sim \beta\,\sqrt{y-y_{\star}}$. Hence 

\begin{equation}
S_{\rm D3-tube}\sim m\,\beta_1\,(y-y_{\star})\,.
\end{equation}
It should be stressed that this result does not rely on a perturbative expansion in $\beta_1$. 

Thus we see that over a distance $y-y_{\star}\sim (m\beta_1)^{-1}$, the contribution of the tube $e^{-S_{\rm D3-tube}}$ is much bigger than the contribution of the geodesic itself, which goes as $e^{-m}$. This shows that there is a ``boundary layer" effect over distances $(m\beta_1)^{-1}$ in which there could be important effects correcting the picture in terms of the geodesic followed by the D3 brane.\footnote{It is straightforward to check that, for standard geodesics, the action for a similar tube action does not lead to a boundary layer.} It is tempting to conjecture that properly taking into account this boundary layer leads to a smoothing of the non-analytic $\Theta(\beta_1)$ behavior. In particular, such behavior might be an artifact of our geodesic approximation, which takes first the $m\rightarrow \infty$ limit --erasing the boundary layer-- and only later $\beta_1\rightarrow 0$. Indeed, as stressed, for the heuristic argument for the existence of the boundary layer we are not relying on a perturbative expansion in $\beta_1$ --while we are implicitly assuming $m$ large by the mere use of a probe brane--, so one could generically estimate the boundary layer length as $m^{-1}$. This might generate a prefactor proportional to $m^{-1}$. On the other hand, the boundary layer thickness is really controlled by $m\beta_1$. We could then imagine, for instance, the double scaling limit where we send $\beta_1\rightarrow 0$ with fixed $m\beta_1$, so that the would-be $m^{-1}$ should be interpreted as $\beta_1/(m\beta_1)\sim \beta_1$, thus suggesting that indeed the $\Theta$ is smoothed.

\section{Conclusions}\label{sec:conclusions}

In this paper we have computed holographically correlation functions for giant graviton operators in $\mathcal{N}=4$ SYM in the presence of monodromy defects. The monodromy defects of interest are the most basic ones one can imagine. They are simply defined by the monodromy $\beta_I$ of charged operators as they go around, with no other data decorating them. This consideration led us to a particular choice of gauge for the SUGRA fields. As we have shown, the computation boils down to the study of charged geodesics hanging from the insertion points of the operators at the boundary in the corresponding 5d gauged SUGRA background. In the presence of the defect, in addition to the standard U-shaped geodesic, there is an extra contribution anchored to the defect. Our final result, to leading order in a perturbative expansion in the $\beta_I$, is \eqref{eq:two-point}. It would be very interesting to perform a conformal block decomposition of this result. To do so unambiguously, one should extend the results in this paper to a more generic kinematical regime, relaxing the condition of $\chi=2$. Nevertheless, despite our simplified kinematics, we could extract in a robust way the one-point function of the $OO^{\dagger}$ operator, which comes entirely from the anchored geodesic. This is to be expected, as in the coincidence limit $X\rightarrow 0$ at fixed $Y$ where the correlator is dominated by such one-point function, the standard geodesic dives less and less into the bulk and consequently knows less and less from the defect. The anchored geodesic, in turn, remains anchored and captures the presence of the defect. Moreover, that the anchored geodesic generates a one-point function for the $OO^{\dagger}$ is completely robust, as it only relies on the existence of the geodesic and the large $y$ asymptotics of  the background. While this is very natural, a puzzling feature of our construction is that the anchored geodesic saddle abruptly disappears when $\beta_1\rightarrow 0$; a kind of non-analytic behavior surely not expected in (finite $N$) QFT. We have proposed a mechanism smoothing such transition through the nucleation of tubes breaking the geodesic controlled by the scale set by $m\beta_1$. It would be extremely interesting to further study this proposal.

It would be interesting to generalize our studies to other dimensions and other backgrounds. Note in particular that, since the anchored geodesic only relies on the existence of the singularity at the defect core and the asymptotics, a one-point function will also be generated. More generically, it would be very interesting to study defects with non-trivial degrees of freedom by allowing $h_I\ne 0$. At first sight, an analytical treatment seems out of reach. It might be however that in an appropriate kinematical regime there is a controlled analytic approach to the problem. In any case, even if having to resort to numerics, going beyond $h_I=0$ and small $\beta_I$ would be extremely interesting.

Finally, regarding the monodromy defect as the endpoint of a topological symmetry defect, it would be very interesting to interpret our results in light of the proposed holographic symmetry operators in \cite{Calvo:2025usj}. In that reference it is argued that non-BPS KK monopoles --non-BPS 7d objects with one transverse Taub-NUT direction decaying into the standard KK monopoles-- are the holographic avatar of R-symmetry operators. To investigate the relation with those it is perhaps simplest to consider the single-charge defect, whose Type IIB uplift is \cite{Bomans:2024vii}

\begin{align}
& ds^2_{IIB}=\sqrt{y\,(y+q_1\,\sin^2\psi)}\Big[ds_{AdS_3}^2+\frac{\sin^2\psi}{y+q_1\,\sin^2\psi}\,ds_{S^3}^2+\frac{dy^2}{4y^2\,(y+q_1-1)}+\frac{y+q_1-1}{y+q_1}d\theta^2\\ \nonumber 
& +\frac{d\psi^2}{y}+\frac{(y+q_1)}{y\,(y+q_1\,\sin^2\psi)}\,\cos^2\psi\,(d\phi-A_{12}\,d\theta)^2\Big]\,.
\end{align}
We can now probe the geometry with a non-BPS KK monopole along $\{AdS_3,S^3,\psi\}$ with Killing direction $\phi$. Following \cite{Calvo:2025usj}, it would couple to

\begin{equation}
d(\imath_k\mathcal{N})\sim q_1\,\frac{y^{\frac{3}{2}}\,\cos\psi\,\sin^3\psi}{(y+q_1)\,\sqrt{y+q_1\sin\psi}}\,{\rm Vol}(AdS_3)\wedge {\rm Vol}(S^3)\wedge d\psi\,.
\end{equation}
Hence, at large $y$, the action for the non-BPS KK monopole would be (we regularize the volume of $AdS$ in the standard way so that we get a finite result)

\begin{equation}
S_{\overline{KK}_B}=\alpha\,(h_1+\beta_1)\,.
\end{equation}
This suggests that indeed the background might contain a non-BPS monopole. In addition, the existence of the geodesic anchored to the singularity at $y_{\star}$ is suggestive of an object sitting at $y_{\star}$ whose degrees of freedom are controlled by $h_I$. It would be very interesting to clarify these issues, as  it would provide a brane realization (in particular including non-BPS KK monopoles) of the monodromy defect. As KK monopoles are arguably more exotic objects than D-branes, it would be very interesting to construct the SUGRA solution for monodromy defects of the baryonic symmetry in the Klebanov-Witten theory (a natural guess is that it should appear as a solution within the truncation in \cite{Cassani:2010na}). In this case the symmetry operators are non-BPS D4 branes \cite{Bergman:2024aly} and the natural correlation function to compute would be that of baryon operators. We postpone these very interesting questions for further studies.

\section*{Acknowledgments}

I would like to thank Y. Lozano for discussions.  This work is supported in part by the Spanish national grant MCIU-25-PID2024-161500NB-I00.

\begin{appendix}

\section{Analyticity of the geodesic action}
\label{app:analyticity}

In this appendix we collect some technical remarks about the evaluation of the integral in \eqref{os}. Since we are doing a perturbative expansion in the $\beta$'s, naively one would expand the integrand, perform the integration, evaluate on the limits (which themselves also depend on the $\beta$'s) and expand again; finding then a non-analytic behavior. To understand this, let us consider the following toy example

\begin{equation}
I=\int_{\beta}^{y_M}\frac{1}{\sqrt{y-\beta}}=2\sqrt{y_M-\beta}\,.
\end{equation}
The exact result admits a completely honest series expansion in $\beta$

\begin{equation}
I=2\sqrt{y_M}-\frac{\beta}{\sqrt{y_M}}+\cdots\,.
\end{equation}

Suppose instead expanding the integrand and then integrating

\begin{equation}
I\sim \int_{\beta}^{y_M} \frac{1}{y^{\frac{1}{2}}}+\frac{\beta}{2y^{\frac{3}{2}}}+\cdots\sim 2\sqrt{y_M}-\sqrt{\beta}-\frac{\beta}{\sqrt{y_M}}+\cdots\,,
\end{equation}
Thus we see that we have produced an artificial $\sqrt{\beta}$ dependence. As anticipated, our case is precisely of this form.\footnote{To make it more explicit, consider redefining $\beta_i\rightarrow \beta\,\beta_i$, so that we are really taking the limit $\beta\rightarrow 0$.} In fact, recall from \eqref{os}

\begin{equation}
S=\int_{y_{\rm m}}^{y_M}dy\,\frac{f_1^2\,f_2+f_3\,(J-f_3)}{\sqrt{f_2}\,\sqrt{f_1^2\,f_2-(J-f_3)^2}}\,.
\end{equation}
where $y_{\rm m}$ --which depends on $\beta$-- is defined as the largest root of $f_1^2\,f_2-(J-f_3)^2$. Thus, we may write

\begin{equation}
f_1^2\,f_2-(J-f_3)^2=(y-y_m)\,F(y)\,,
\end{equation}
so

\begin{equation}
S=\int_{y_{\rm m}}^{y_M}dy\, \frac{G(y;\beta)}{\sqrt{y-y_m}}\,,\qquad G(y;\beta)=\frac{f_1^2\,f_2-f_3\,(J-f_3)}{\sqrt{f_2}\,\sqrt{F}}\,.
\end{equation}
Note that $G(y_{\rm m};\beta)$ is some finite number. Since we are interested on the lower part of the integral, let us write $y=x+y_{\rm m}$, so that

\begin{equation}
S=\int_{0}^{y_M-y_{\rm m}}dx\, \frac{G(x+y_{\rm m};\beta)}{\sqrt{x}}\,.
\end{equation}
From this point of view, clearly the lower end of the integral does not contribute any $\beta$ dependence: everything comes from the region near the upper end $y_M-y_{\rm m}$. We can now safely expand in $\beta$ (we write $y_{\rm m}=y_{\rm m}^{(0)}+\beta\,y_{\rm m}^{(1)}+\cdots$). Schematically

\begin{equation}
\frac{G(x+y_{\rm m};\beta)}{\sqrt{x}}=\frac{G_0(x)}{\sqrt{x}}+\beta\,\frac{G_1(x)}{\sqrt{x}}+\cdots\,,
\end{equation}
so the integral is

\begin{equation}
S= \Big[\int_{0}^{y_M-y_{\rm m}^{(0)}}\frac{G_0(x)}{\sqrt{x}}\Big] - \beta\,\Big[ y_{\rm m}^{(1)}\, \frac{G_0(y_M-y_{\rm m}^{(0)})}{\sqrt{y_M-y_{\rm m}^{(0)}}}-\,\int_{0}^{y_M-y_{\rm m}^{(0)}}\frac{G_1(x)}{\sqrt{x}}\Big] +\cdots \,.
\end{equation}
In particular, we see that the $\beta$ expansion is perfectly analytic in $\beta$. Reassuringly, applying this to our toy model, where $G=1$ --so $G_0=1$ and $G_1=0$--, $y_{\rm m}^{(0)}=0$, $y_{\rm m}^{(1)}=1$; recovers the correct result.

\end{appendix}


\begin{thebibliography}{99}

\bibitem{Calvo:2025usj}
H.~Calvo, F.~Mignosa and D.~Rodriguez-Gomez,
``R-symmetries, anomalies and non-invertible defects from non-BPS branes,''
JHEP \textbf{10} (2025), 107
doi:10.1007/JHEP10(2025)107
[arXiv:2506.13859 [hep-th]].

\bibitem{Bah:2025vfu}
I.~Bah, F.~Bonetti, M.~Chitoto and E.~Leung,
``Non-Abelian Symmetry Operators from Hanging Branes in $AdS_5 \times S^5$,''
[arXiv:2510.19812 [hep-th]].

\bibitem{Balasubramanian:2001nh}
V.~Balasubramanian, M.~Berkooz, A.~Naqvi and M.~J.~Strassler,
``Giant gravitons in conformal field theory,''
JHEP \textbf{04} (2002), 034
doi:10.1088/1126-6708/2002/04/034
[arXiv:hep-th/0107119 [hep-th]].

\bibitem{McGreevy:2000cw}
J.~McGreevy, L.~Susskind and N.~Toumbas,
``Invasion of the giant gravitons from Anti-de Sitter space,''
JHEP \textbf{06} (2000), 008
doi:10.1088/1126-6708/2000/06/008
[arXiv:hep-th/0003075 [hep-th]].

\bibitem{Grisaru:2000zn}
M.~T.~Grisaru, R.~C.~Myers and O.~Tafjord,
``SUSY and goliath,''
JHEP \textbf{08} (2000), 040
doi:10.1088/1126-6708/2000/08/040
[arXiv:hep-th/0008015 [hep-th]].

\bibitem{Billo:2016cpy}
M.~Bill{\`o}, V.~Gon{\c{c}}alves, E.~Lauria and M.~Meineri,
``Defects in conformal field theory,''
JHEP \textbf{04} (2016), 091
doi:10.1007/JHEP04(2016)091
[arXiv:1601.02883 [hep-th]].

\bibitem{Cvetic:2000nc}
M.~Cvetic, H.~Lu, C.~N.~Pope, A.~Sadrzadeh and T.~A.~Tran,
``Consistent SO(6) reduction of type IIB supergravity on S**5,''
Nucl. Phys. B \textbf{586} (2000), 275-286
doi:10.1016/S0550-3213(00)00372-2
[arXiv:hep-th/0003103 [hep-th]].

\bibitem{Kunduri:2007qy}
H.~K.~Kunduri and J.~Lucietti,
``Near-horizon geometries of supersymmetric AdS(5) black holes,''
JHEP \textbf{12} (2007), 015
doi:10.1088/1126-6708/2007/12/015
[arXiv:0708.3695 [hep-th]].

\bibitem{Ferrero:2021etw}
P.~Ferrero, J.~P.~Gauntlett and J.~Sparks,
``Supersymmetric spindles,''
JHEP \textbf{01} (2022), 102
doi:10.1007/JHEP01(2022)102
[arXiv:2112.01543 [hep-th]].

\bibitem{Arav:2024exg}
I.~Arav, J.~P.~Gauntlett, Y.~Jiao, M.~M.~Roberts and C.~Rosen,
``Superconformal monodromy defects in $ \mathcal{N} $=4 SYM and LS theory,''
JHEP \textbf{08} (2024), 177
doi:10.1007/JHEP08(2024)177
[arXiv:2405.06014 [hep-th]].

\bibitem{Bomans:2024vii}
P.~Bomans and L.~Tranchedone,
``Holographic generalised Gukov-Witten defects,''
JHEP \textbf{03} (2025), 118
doi:10.1007/JHEP03(2025)118
[arXiv:2410.18172 [hep-th]].

\bibitem{Conti:2025wyj}
A.~Conti, Y.~Lozano and C.~Rosen,
``Monodromy Defects in Massive Type IIA,''
[arXiv:2512.10006 [hep-th]].

\bibitem{Gomis:2007fi}
J.~Gomis and S.~Matsuura,
``Bubbling surface operators and S-duality,''
JHEP \textbf{06} (2007), 025
doi:10.1088/1126-6708/2007/06/025
[arXiv:0704.1657 [hep-th]].

\bibitem{Choi:2024ktc}
C.~Choi, J.~Gomis and R.~Izquierdo Garc{\'\i}a,
``Surface operators and exact holography,''
JHEP \textbf{12} (2024), 195
doi:10.1007/JHEP12(2024)195
[arXiv:2406.08541 [hep-th]].

\bibitem{IzquierdoGarcia:2025jyb}
R.~Izquierdo Garc{\i}a,
[arXiv:2512.12696 [hep-th]].

\bibitem{Gukov:2006jk}
S.~Gukov and E.~Witten,
``Gauge Theory, Ramification, And The Geometric Langlands Program,''
[arXiv:hep-th/0612073 [hep-th]].

\bibitem{Janik:2010gc}
R.~A.~Janik, P.~Surowka and A.~Wereszczynski,
``On correlation functions of operators dual to classical spinning string states,''
JHEP \textbf{05} (2010), 030
doi:10.1007/JHEP05(2010)030
[arXiv:1002.4613 [hep-th]].

\bibitem{Bak:2011yy}
D.~Bak, B.~Chen and J.~B.~Wu,
``Holographic Correlation Functions for Open Strings and Branes,''
JHEP \textbf{06} (2011), 014
doi:10.1007/JHEP06(2011)014
[arXiv:1103.2024 [hep-th]].

\bibitem{Bissi:2011dc}
A.~Bissi, C.~Kristjansen, D.~Young and K.~Zoubos,
``Holographic three-point functions of giant gravitons,''
JHEP \textbf{06} (2011), 085
doi:10.1007/JHEP06(2011)085
[arXiv:1103.4079 [hep-th]].

\bibitem{Yang:2021kot}
P.~Yang, Y.~Jiang, S.~Komatsu and J.~B.~Wu,
``D-branes and orbit average,''
SciPost Phys. \textbf{12} (2022) no.2, 055
doi:10.21468/SciPostPhys.12.2.055
[arXiv:2103.16580 [hep-th]].

\bibitem{Holguin:2025dei}
A.~Holguin,
``Semiclassics, branes, and extremality,''
[arXiv:2512.24979 [hep-th]].

\bibitem{Bianchi:2021snj}
L.~Bianchi, A.~Chalabi, V.~Proch{\'a}zka, B.~Robinson and J.~Sisti,
``Monodromy defects in free field theories,''
JHEP \textbf{08} (2021), 013
doi:10.1007/JHEP08(2021)013
[arXiv:2104.01220 [hep-th]].

\bibitem{Cassani:2010na}
D.~Cassani and A.~F.~Faedo,
``A Supersymmetric consistent truncation for conifold solutions,''
Nucl. Phys. B \textbf{843} (2011), 455-484
doi:10.1016/j.nuclphysb.2010.10.010
[arXiv:1008.0883 [hep-th]].

\bibitem{Bergman:2024aly}
O.~Bergman, E.~Garcia-Valdecasas, F.~Mignosa and D.~Rodriguez-Gomez,
``Non-BPS branes and continuous symmetries,''
JHEP \textbf{02} (2025), 066
doi:10.1007/JHEP02(2025)066
[arXiv:2407.00773 [hep-th]].





\end{thebibliography}
\end{document}